\documentclass[letterpaper,12pt]{article}
\usepackage{amssymb}
\usepackage{dsfont}
\usepackage{amsmath}
\usepackage{amsthm}
\usepackage{lineno}
\usepackage{epsfig}
\usepackage{graphics}
\usepackage{graphicx}
\usepackage{color}
\usepackage[nofiglist,notablist]{endfloat}
\usepackage{subfig}
\usepackage{makecell}
\usepackage{colortbl}  
\usepackage{xcolor}
\usepackage{array}
\usepackage{threeparttable}
\usepackage{booktabs}
\usepackage[authoryear]{natbib}
\usepackage[colorlinks,linkcolor=blue,anchorcolor=blue,citecolor=blue,]{hyperref}
\usepackage[top=1in,bottom=1in,left=1.1in,right=1.1in]{geometry}
\usepackage[capitalize]{cleveref}
\usepackage{caption}
\usepackage{abstract}
\usepackage{endnotes}
\usepackage{setspace}

\usepackage{authblk}
\begin{document}

\begin{spacing}{1.5} 

\title{ \Large 
Enhancing Black-Scholes Delta Hedging via Deep Learning}%
\author{
Chunhui Qiao
and 
Xiangwei Wan
\thanks{Antai College of Economics and Management, Shanghai Jiao Tong University, Shanghai, China. 
Emails:  chunhuiqiao@sjtu.edu.cn and xwwan@sjtu.edu.cn} 
}

\date{\today
}
\maketitle
\vspace{-0.2in}

\begin{abstract} 

This paper proposes a deep delta hedging framework for options, utilizing neural networks to learn the residuals between the hedging function and the implied Black-Scholes delta. 
This approach leverages the smoother properties of these residuals, enhancing deep learning performance.
Utilizing ten years of daily S\&P 500 index option data, our empirical analysis demonstrates that learning the residuals
, using the mean squared one-step hedging error as the loss function, significantly improves hedging performance over directly learning the hedging function, often by more than 100\%. 
Adding input features when learning the residuals enhances hedging performance more for puts than calls, with market sentiment being less crucial. 
Furthermore, learning the residuals with three years of data matches the hedging performance of directly learning with ten years of data, 
proving that our method demands less data.

~

\end{abstract}

\end{spacing}

\newpage

\section{Introduction}  

Delta hedging is a fundamental risk management strategy in options trading, often implemented using the Black-Scholes model. Traders dynamically adjust their hedging positions on the underlying asset according to the Black-Scholes delta, which is the partial derivative of the Black-Scholes option valuation function with respect to the price of the underlying asset. In practice, the implied Black-Scholes delta, also called the practitioner Black-Scholes delta, derived from market prices of options, is frequently used due to its incorporation of current market conditions. However, model-based approaches like Black-Scholes delta hedging or its extensions (e.g., \citealp{bakshi1997})
have inherent limitations, such as model misspecification, which can lead to significant hedging errors.

In contrast to these traditional methods, data-driven approaches offer a promising alternative for option hedging. One such approach is based on minimizing the mean squared of the one-step hedging error in a discrete setting. By framing the hedging problem as a data-driven optimization task, deep learning can be used to learn the hedging function directly from the data (e.g., \citealp{Nian2021}, \citealp{Ruf2022}, \citealp{Chen2023}). While data-driven approaches are robust to specification errors, they often lack interpretability and are highly data-intensive, requiring large quantities of historical prices to obtain a sufficiently well-trained network.

In this paper, we follow the routine of the data-driven approach while retaining the economic structure of the implied Black-Scholes delta.
\footnote{\cite{chen2024teaching} propose a novel framework that combines economic restrictions from structural models with machine learning through transfer learning, significantly improving model performance, especially in option pricing with limited data or volatile markets.} 
Specifically, we utilize neural networks to learn the residuals between the hedging function and the implied Black-Scholes delta from option data, using the mean squared of the one-step hedging error as the loss function.

Several technical and economic reasons motivate our approach of learning the residuals rather than directly learning the hedging function. 
Technically, the former is likely to be smoother than the latter, which can potentially be approximated by a simpler neural network and demand less data for learning.
Even the universal approximation theorem (see, e.g., \citealp{cybenko1989}; \citealp{hornik1989}) ensures that deep neural networks can approximate any continuous function, a function with better smoothness properties can be approximated faster with less complex networks (see, e.g. \citealp{barron1993}). For options nearing expiration, their hedging function is discontinuous, as the optimal hedging strategy would be either a full hedge or no hedge, depending on whether the option is in-the-money or out-of-the-money.  The implied Black-Scholes delta shares this discontinuity. The difference between  the hedging function and the implied Black-Scholes delta will eliminate this discontinuity, making it more conducive to learning by neural networks. 

Economically, the implied Black-Scholes delta is a direct measure of the sensitivity of the option's price to changes in the price of the underlying asset, incorporating the updated market expectation for volatility. 
Extended models also consider the indirect effects of changes in the underlying asset's price on the option's price as corrections to the implied Black-Scholes delta. For example, some literature corrects  the implied Black-Scholes delta with additional terms that account for changes in implied volatility (e.g., \citealp{crepey2004}, \citealp{bartlett2006}, \citealp{alexander2012}, \citealp{Hull2017}). We extend this framework by using deep learning to learn the correction terms, utilizing the mean squared of the one-step hedging error as the loss function.

We hypothesize that this hybrid approach, combining the strengths of model-based and data-driven strategies, can mitigate the shortcomings of each method and significantly enhance hedging performance. To validate this hypothesis, we conducted extensive empirical analysis using ten years of S\&P 500 index option data. We specifically opted for Feedforward Neural Networks (FNN) for deep learning, as complex neural networks have the potential to extract more information from data but may generate a large number of parameters, leading to overfitting.

In our analysis, we utilized common option features, including time to maturity, moneyness, implied volatility, and option Greeks such as implied Black-Scholes delta, theta, vega, and gamma (see also \citealp{Nian2021}). Additionally, we incorporated the sentiment feature from \cite{Chen2023}, which includes the volatility index (VIX) for calls and index return for puts. The empirical results validate the effectiveness of our approach. Specially, our key findings include: 
(1) Hedging performance from learning the residuals significantly surpasses that from directly learning the hedging function, often showing improvements exceeding 100\%. 
(2) When learning the residuals, adding input features improves daily hedging performance more effectively for puts than for calls. Additionally, for both calls and puts, the market sentiment variable is not as crucial as it is when directly learning the hedging function. 
(3) Learning the residuals with just three years of data can achieve hedging performance comparable to those obtained by directly learning the hedging function with ten years of data.
 
    \subsection{Related literature} \label{sec:literature}

Since \cite{Hutchinson1994}  proposed the use of non-parametric models to estimate option pricing and hedging problems, numerous neural networks and machine learning methods have been proposed and applied to option pricing. For instance, \cite{garcia2000}, \cite{bennell2004}, \cite{Gradojevic2009} have explored various approaches in this domain. In contrast to \cite{Hutchinson1994} , who advocated learning option prices and then deriving the delta hedging function through differentiation, \cite{Hull2017} and \cite{Nian2018} have employed data-driven methods to directly fit the delta hedging function, aiming to minimize local hedging errors.
\cite{Hull2017} utilized a quadratic parameterization model concerning the implied Black-Scholes delta, while \cite{Nian2018} treated the delta hedging function as a kernel function of features such as time to maturity, moneyness, and the implied Black-Scholes delta, employing spline function fitting for hedging. Their studies demonstrate that the data-driven delta hedging model outperforms the parametric model of \cite{Hull2017}  and other model-based approaches, including the local volatility models of \cite{derman1994} and \cite{dupire1994}, as well as the stochastic volatility models (SABR) of  \cite{hagan2002}.

\cite{Nian2021} extend the work of \cite{Nian2018} by introducing  a robust encoder-decoder Gated Recurrent Unit (GRU) model, which integrates both local and time-sequential features to learn the hedging function. 
Sequential features, such as historical implied volatility and Black-Scholes delta, are input into the GRU to generate a hidden feature. 
By combining this hidden feature with local features, two FNNs generate a candidate hedge ratio and a weight function.  Subsequently, the final hedging function is derived as the weighted average of the candidate hedge ratio and the implied Black-Scholes delta. Their numerical experiments
demonstrate that  this hedging function outperforms  the kernel model by \cite{Nian2018}, the parametric model of \cite{Hull2017}  and  the SABR model of  \cite{hagan2002}. In comparison, we do not pursue the use of more complex deep learning methods. Instead, we leverage the smoother functional properties of the residual function, allowing it to be learned using a standard FNN. Our empirical results using three years of data show that the gain ratios from learning the residual function via a standard FNN are even better  those obtained in \cite{Nian2021} using more complex neural network models.

\cite{Ruf2022} utilized a  FNN and linear regression to fit the hedging function. Specifically, they emphasized the simplicity of linear regression methods, highlighting that including the implied Black-Scholes delta as an independent variable can yield hedging performance similar to that of the FNN method. 
\cite{Chen2023} further investigated the empirical performance of FNN methods in learning the hedging function, focusing on the influence of the size of the training data and the selection of input features. 
 They concluded that increasing the volume of data, from the three-year dataset used by \cite{Nian2021} and \cite{Ruf2022} to a decade's worth, and considering market sentiment indicators, such as the volatility index for calls and the index return for puts, significantly improved the local hedging effectiveness of FNNs. 
 They showed that deep learning can significantly outperform the  model in \cite{Hull2017} in the out-of-sample test.
 \cite{Chen2023} also explored using GRU to uncover information from historical VIX and index return data, but their performance did not surpass that of directly using FNNs. In comparison, our study employs FNNs to learn the residuals between the hedging function and the implied Black-Scholes delta. Our empirical results demonstrate that this approach leads to a significant improvement in hedging performance compared to directly learning the hedging function, often exceeding 100\%. 
 In our approach, the market sentiment variable is less crucial than when directly learning the hedging function. 
 Furthermore, learning the residuals with three years of data achieves hedging performance comparable to directly learning the hedging function with ten years of data.

The optimal local hedging strategies considered in the aforementioned literature, including this paper, focus on an option trader's optimal delta hedging strategy when selling an option at the beginning and buying it back at the end in a short single period, as discussed in \cite{bergomi2015}. Another class of strategies in the literature addresses minimizing global hedging errors. A notable example is \cite{Buehler2019}, which employs deep learning to learn the optimal delta hedging strategy that minimizes a convex risk measure over the hedging errors at option expiration. Research on global hedging strategies often involves using reinforcement learning methods to optimize dynamic hedging strategies. Relevant studies include 
\cite{halperin2019qlbs,halperin2020qlbs}, \cite{kolm2019}, \cite{du2020}, \cite{cao2021deep}, \cite{Zhang2021},  \cite{carbonneau2021}, \cite{Dai2023}  and \cite{mikkila2023}.
 
Another  strands of literature  investigates using neural networks to fit and predict the movement of implied volatility, see, e.g., \cite{Hull2020} and  \cite{Zhang2023}. As mentioned in \cite{Hull2017}, the optimal delta hedging function in a local risk minimization problem is the implied Black–Scholes delta plus the implied Black–Scholes vega times the partial derivative of the expected implied volatility with respect to the asset price (see also \citealp{crepey2004}, \citealp{bartlett2006}, \citealp{alexander2012}). Once the movement of implied volatility is learned, a correction to the implied Black-Scholes delta can be obtained. Unlike \cite{Hull2020}, in this paper, we use a neural network with the objective of directly minimizing the mean squared of the one-step hedging error.

It is also worth mentioning the works of \cite{fu2022-QF, fu2022-WP}. They use deep learning to solve partial differential equations for pricing standard European options and barrier options. 
Recognizing that option payoffs are non-smooth at the strike and barrier levels, they add singular terms to the neural networks, enabling the networks to replicate the asymptotic behaviors of option payoffs at short maturities. Similarly, this paper addresses the issue of the non-smoothness of the hedging function but employs a simpler method by learning the residuals to eliminate singularities. 


	The rest of the paper is organized as follows. \cref{sec2} explains the descriptive statistics of data, \cref{sec3} presents the deep learning models, 
 \cref{sec4} shows the empirical results. The concluding remarks are given in \cref{sec5}.

\section{Data}\label{sec2}
	Our data comes from the OptionMetrics. We utilized ten years of data on S\&P 500 index options, which are  European-style options, covering the period from January 1, 2010, to December 31, 2019. This database provided closing bid and ask prices for the option contracts along with option sensitivities such as the implied Black-Scholes delta ($\delta_{BS}$), theta ($\theta_{BS}$), vega ($v_{BS}$), gamma ($\gamma_{BS}$) and others. We followed the data filtering approach outlined by \cite{Hull2017}, excluding option quotes that were untraded or had missing information on bid prices, ask prices, implied volatility, $\delta_{BS}$, $\theta_{BS}$, $v_{BS}$, or $\gamma_{BS}$.  Additionally, we removed options with time to maturity less than 14 days, and chose call options with $\delta_{BS}$ in the range [0.05, 0.95] and put options with $\delta_{BS}$ in the range [-0.95, -0.05], as deeply out-of-the-money and deeply in-the-money options, as well as those with very short expiration dates, tend to be noisy and unreliable.

 After all the filtering, we were left with more than 2.073 million price quotations for both puts and calls on the S\&P 500. A summary of the statistical properties of these options is provided in \cref{data}. The data shows that most options are concentrated in at-the-money and out-of-the-money categories.
	\begin{center}
		\begin{table}[!htb]
			\caption{Percentage of trading volume and the number for each delta bucket of S\&P 500 index options between Jan 1, 2010 and Dec 31, 2019.}\label{data}
			\scalebox{0.8}{
				\tabcolsep=0.85cm
				\begin{tabular}{llrllr}				
					\hline
					Delta bucket&Call&number&Delta bucket&Put&number\\
					\hline
					0.1&0.1929&177397&-0.1&0.2989&427684\\
					0.2&0.1628&113802&-0.2&0.2093&244134\\
					0.3&0.1328&98583&-0.3&0.1410&167260\\
					0.4&0.1248&97994&-0.4&0.1335&134925\\
					0.5&0.2701&115378&-0.5&0.1734&109754\\
					0.6&0.0776&82713&-0.6&0.0277&61063\\
					0.7&0.0212&66088&-0.7&0.0092&41028\\
					0.8&0.0107&50210&-0.8&0.0044&27814\\
					0.9&0.0071&36541&-0.9&0.0025&21297\\
					\hline
     \multicolumn{6}{l}{Note: The delta bucket $d$ contains options with delta in $[d-0.05,d+0.05)$.}
			\end{tabular}}
		\end{table}
	\end{center}
	
	\section{Deep learning model}\label{sec3} 

 \subsection{Learning objective}
In this paper, we learn a hedging strategy by minimizing the mean squared local hedging error. Specially, consider an option trader who holds a portfolio consisting of  one share of option $i$ and $\delta^{(i)}$ shares of underlying asset on the opposite direction. After one period, such as one day, one week, or one month, the trader liquidates the portfolio.
The profit/loss of the portfolio, also known as the local hedging error, is given by
\begin{align*}
		\Delta V_i-\delta^{(i)}\Delta S_i,
\end{align*}
where $\Delta V_i$ and $\Delta S_i$ are the price changes of the option and the underlying asset during this period, respectively. The local delta hedging  problem is to find the optimal function $\delta^{(i)}$ that minimizes the mean squared  local hedging error, expressed as: 
	\begin{align} \label{originallearning}
		\min_{\delta^{(i)}} \frac{1}{M_1}\sum_{i=1}^{M_1}\left(\Delta V_i-\delta^{(i)}\Delta S_i\right)^2,  
	\end{align}
where $M_1$ is the number of option in the training set, as also seen in \cite{Hull2017}, \cite{Nian2018,Nian2021} and \cite{Chen2023}. 

\begin{figure}[!htb]
		\centering
  	\subfloat[]{\includegraphics[width=8cm, height=6cm]{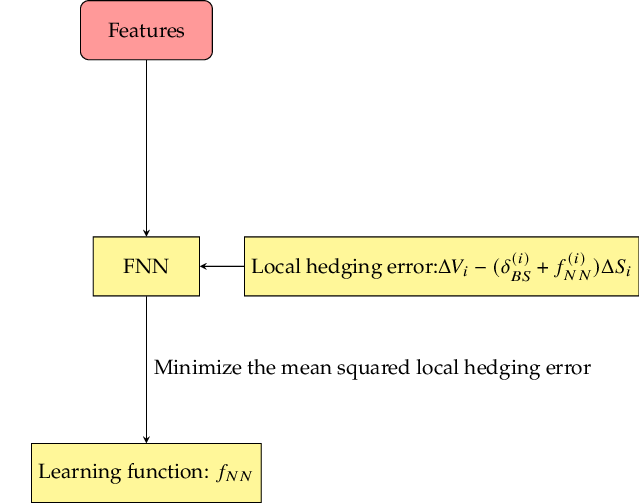}}
        \subfloat[]{\includegraphics[width=8cm, height=6cm]{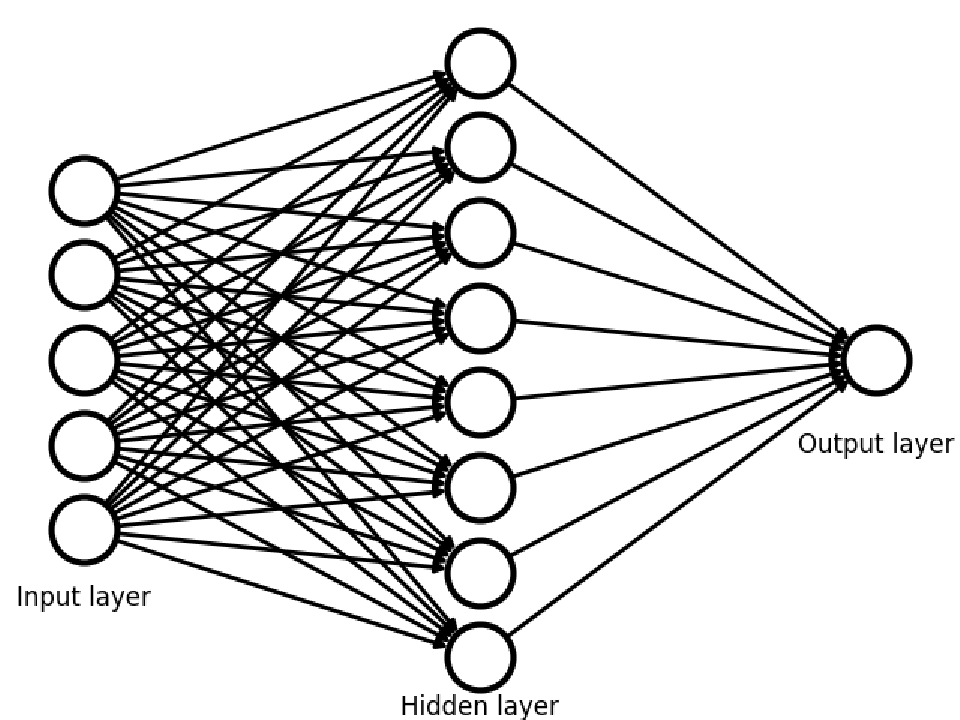}}
		\caption{\footnotesize (a) Workflow of the hedging network.  (b) Structure of one hidden layer FNN. }
		\label{fig_process}
	\end{figure}

For options near maturity, the option prices approximate their payoffs, whose derivative is discontinuous at the strike price. 
 Consequently, the function $\delta^{(i)}$  would approximate the derivative of the option payoff, inheriting the discontinuity and making it difficult to be approximated by commonly used neural networks. Since the implied Black-Scholes delta of  the $i$-th option,  $\delta_{BS}^{(i)}$, shares the same discontinuity, the difference  between  $\delta^{(i)}$  and $\delta_{BS}^{(i)}$ would be smoother and better be approximated by neural networks. Inspired by this intuition, we replace the function $\delta^{(i)}$ in \eqref{originallearning} with $\delta_{BS}^{(i)}+f_{NN}^{(i)}(x)$ to generate our approach to the local delta hedging  problem as follows: 
	\begin{align}\label{residuallearning}
		\min_{f_{NN}^{(i)}(x)}\frac{1}{M_1}\sum_{i=1}^{M_1}\left(\Delta V_i-(\delta_{BS}^{(i)}+f_{NN}^{(i)}(x))\Delta S_i\right)^2,
	\end{align}
	where $f_{NN}^i(x)$ is the residual of the $i$-th option's hedging position, which will be learned by an FNN described in next subsection, and $x$ is the input features of the neural networks. 
 The overall process of our heading model is shown in the \cref{fig_process}(a).
 
	\subsection{The structure of neural networks}
 In this paper, we use the  Feedforward Neural Networks (FNN),
  also known as the Multi-Layer Perceptron (MLP), to learn the residual $f_{NN}^i(x)$ in \eqref{residuallearning}. The structure of an FNN with one hidden layer is shown in \cref{fig_process} (b).  
 
 Let $\textbf{X}\in \mathbb{R}^{d}$ be the vector of input features, where $d$ represents the number of input features, and $\textbf{O}\in \mathbb{R}$ be the final output layer. For a one-hidden-layer FNN, let $\textbf{X}^{(1)}\in \mathbb{R}^{h}$ be the hidden layer, where $h$ represents the number of neurons 
 on the hidden layer. The hidden layer and output layer are fully connected, with
 hidden layer weights $\textbf{W}^{(1)}\in \mathbb{R}^{d\times h}$ and bias $\textbf{b}^{(1)}\in \mathbb{R}^{1\times h}$, 
 output layer weights $\textbf{W}^{(2)}\in \mathbb{R}^{h}$ and bias $\textbf{b}^{(2)}\in \mathbb{R}$. 
 Using an activation function $\sigma(\cdot)$, the input-to-output mapping is calculated as follows:
	\begin{align*}
		\textbf{X}^{(1)} &= \sigma(\textbf{X}\textbf{W}^{(1)} + \textbf{b}^{(1)}),\\
		\textbf{O} &= \textbf{X}^{(1)}\textbf{W}^{(2)} + \textbf{b}^{(2)}.
	\end{align*}
	Similarly, for an FNN with $N$ hidden layers, the input-to-output mapping is given as follows:
	\begin{align*}
        \textbf{X}^{(1)} &= \sigma(\textbf{X}\textbf{W}^{(1)} + \textbf{b}^{(1)}),\\
		\textbf{X}^{(i+1)} &= \sigma(\textbf{X}^{(i)}\textbf{W}^{(i+1)}+\textbf{b}^{(i+1)}), i=1,...,N-1,  \\
		\textbf{O} &= \textbf{X}^{(N)}\textbf{W}^{(N+1)} + \textbf{b}^{(N+1)}.
	\end{align*}
In our empirical study, when using the full ten-year dataset, our neural network employs 3 hidden layers with 128 neurons each, i.e., $N=3$ and $h=128$. When using a subset of three-year data, our neural network employs 2 hidden layers with 128 neurons each, i.e., $N=2$ and $h=128$. For the activation function, we use the sigmoid function, that is, $\sigma(x)=1/(1+\exp(-x))$.

 \subsection{Features selection}

 Selecting features that significantly impact the hedging function is both an important and interesting question. Since this paper focuses on investigating whether learning the residuals, as opposed to directly learning the hedging function, can improve hedging performance, we have chosen some commonly used option characteristics and option Greeks for our analysis. The option characteristics include time to maturity (TTM), moneyness, and implied volatility ($\sigma_{imp}$). The option Greeks include the implied Black-Scholes delta ($\delta_{BS}$), theta ($\theta_{BS}$), vega ($v_{BS}$), and gamma ($\gamma_{BS}$). 
In our empirical studies, we test the hedging performance of both learning the hedging function and learning the residuals for different sets of features. These feature sets are summarized in \cref{features}. \cite{Chen2023} demonstrates that learning the hedging function using an FNN model with three features (TTM, $\delta_{BS}$, and VIX for calls/index return for puts
performs best overall among alternatives and also outperforms the model proposed by \cite{Hull2017}. We also use the features from \cite{Chen2023}’s three-feature model, referred to as Fea3-CL in \cref{features}. Note that in \cref{features}, the model names in the first column appended with "-BS" indicate learning the residuals, while those without "-BS" indicate directly learning the hedging function.

	\begin{center}
		\begin{table}[!htb]
			\caption{A summary of various types of deep learning models}\label{features}
			\scalebox{0.8}{
				\tabcolsep=1.3cm
				\begin{tabular}{ll} 		
					\hline
					Models&\multicolumn{1}{l}{Features} \\ 
					\hline
					Fea3-CL/Fea3-CL-BS&TTM, $\delta_{BS}$, VIX for calls/index return for puts\\
					Fea2/Fea2-BS&TTM, $\delta_{BS}$\\
					Fea3/Fea3-BS&TTM, $\delta_{BS}$, Moneyness\\
					Fea4/Fea4-BS&TTM, $\delta_{BS}$, Moneyness, $\sigma_{imp}$\\
					Fea5/Fea5-BS&TTM, $\delta_{BS}$, Moneyness, $\sigma_{imp}$, $\theta_{BS}$\\
					Fea6/Fea6-BS&TTM, $\delta_{BS}$, Moneyness, $\sigma_{imp}$, $\theta_{BS}$, $v_{BS}$\\
					Fea7/Fea7-BS&TTM, $\delta_{BS}$, Moneyness, $\sigma_{imp}$, $\theta_{BS}$, $v_{BS}$, $\gamma_{BS}$\\
					\hline
					\multicolumn{2}{l}{Note: Model names in the first column appended with ”-BS” indicate learning the residuals, }\\
     \multicolumn{2}{l}{while those without ”-BS” indicate directly learning the hedging function.}
			\end{tabular}}
		\end{table}
	\end{center}
 
 \subsection{Hedging performance criteria}
	 To compare each approach, we follow \cite{Hull2017} to use the Gain ratio as the hedging performance evaluation criteria. Specially, for the hedging function $\delta$ learned directly from \eqref{originallearning} and the residual function $f_{NN}$ learned from \eqref{residuallearning}, the Gain ratio are computed by 
  \begin{align}\label{gain-o}
		\mathrm{Gain \ Ratio} (\delta)=1-\frac{\sum_{i=1}^{M_2}(\Delta V_i-\delta^{(i)}\Delta S_i)^2}{\sum_{i=1}^{M_2}(\Delta V_i-\delta^{(i)}_{BS}\Delta S_i)^2},
	\end{align}
 and
	\begin{align}\label{gain-r}
		\mathrm{Gain \ Ratio} (f_{NN})=1-\frac{\sum_{i=1}^{M_2}(\Delta V_i-(\delta_{BS}^{(i)}+f_{NN}^{(i)}(x))\Delta S_i)^2}{\sum_{i=1}^{M_2}(\Delta V_i-\delta^{(i)}_{BS}\Delta S_i)^2},
	\end{align}
 respectively, where $M_2$ is the number of option in the test set. This criterion uses the mean squared local hedging error with the implied Black-Scholes delta as a benchmark, providing an assessment of the improvement in the mean squared local hedging error achieved by the new hedging strategy compared to the benchmark.
 
	\section{Empirical results}\label{sec4}
	\subsection{Data splitting}
In our empirical study, we first use the entire 10-year dataset to test the hedging performance of different models listed in \cref{features} for daily, weekly, and monthly hedging. 
To evaluate whether the residuals can be effectively learned by a simpler neural network with less data, we also tested  the hedging performance of our approach using 3-year subdatasets. The empirical results for this test are presented at the end of this section. 

When using the 10-year dataset, we divided the data into two parts: the first nine years serve as the training and validation set, while the last year serves as the test set (see also \citealp{Hull2020} and \citealp{Chen2023}). We randomly split the data from the first nine years into two parts, with 80\% used as the training set and 20\% as the validation set. The number of observations in the training, validation, and test sets are 531523, 132880, and 174303 for calls, and 780793, 195198, and 258968 for puts, respectively.

	\subsection{Training method}
	
	To mitigate the risk of overfitting, we employed specific criteria to govern the training process. Typically, training was halted when the mean squared error of the validation set begins to escalate. However, in scenarios with substantial datasets, such a turning point often materializes after a prolonged period. 
 In our experiment, we employed fewer epochs than the previous model because the residual function we aimed to learn is smoother and easier for neural networks to approximate.
 
	1. \textbf{Xavier Initialization:}  This method is commonly used in deep neural networks and effectively helps alleviate the problem of vanishing or exploding gradients during training. Specifically, Xavier initialization sets the weights according to the type of activation function (such as ReLU, sigmoid, etc.) used by the input and output layers.
 
	2. \textbf{Gradient Clipping:} This technique used to prevent the problem of gradient explosion during the training of deep neural networks. The principle of gradient clipping is to clip the gradients to a specified range before each weight update, thereby avoiding excessive weight updates caused by overly large gradients. This range is usually determined empirically or experimentally and is typically a relatively small value.
 
	3. \textbf{Batch Normalization:} This special layer in neural networks addresses the problem of internal covariate shift during training, which can accelerate training and improve model generalization. Specifically, Batch Normalization normalizes each batch of data, standardizing the intermediate outputs of the neural network so that each layer's output maintains an appropriate scale, thus avoiding the problem of vanishing or exploding gradients.
 
	We experimented with different hyperparameters and model hidden layer sizes related to training. Ultimately, we employed three hidden layers with 128 neurons each. We set the batch size to 1024 and used the ADAM optimizer. To obtain the optimal solution, we set the learning rate to 0.0001. In our experiments, training typically stopped after about 40 epochs (one epoch consists of the iterations needed to go through all the mini-batches).
 
	\subsection{Daily hedging performance of various models using 10-year data}
 For the different feature sets listed in \cref{features}, we used daily data from the first nine years to learn the residual function $f_{NN}$ by solving optimization problem \eqref{residuallearning} and the hedging function $\delta$ by solving optimization problem \eqref{originallearning}. We then evaluated the hedging performance of each model by computing the Gain Ratio on the test dataset from the final year using equations \eqref{gain-r} and \eqref{gain-o}, respectively. 
  
 
  \cref{tab-daycall} and \cref{tab-dayput} show the gain ratio of each model for daily hedging using calls and puts from the overall test set, as well as from each delta bucket, which is a measure of moneyness (see \citealp{Hull2017}).
	\begin{center}
		\begin{table}[!htb]
			\caption{Gain ratios of daily hedging for S\&P 500 call options in various models}\label{tab-daycall}
			\scalebox{0.8}{
				\tabcolsep=0.4cm
				\begin{tabular}{lllllllll}				
					\hline
					delta bucket&Fea2&\textbf{Fea2-BS}&Fea3&\textbf{Fea3-BS}&Fea4&\textbf{Fea4-BS}&Fea5&\textbf{Fea5-BS}\\ 
					\hline
					0.1&0.2916&	0.6637&0.2604&	0.6626&0.3099&0.6693&0.3173&	0.6793	\\
					0.2&0.2917&	0.6518&0.2359&	0.6385&0.2355&0.6569&0.2245&	0.6686	\\
					0.3&0.1968&	0.5985&0.1971&	0.5919&0.1980&0.6378&0.1992&	0.5976	\\
					0.4&0.3123&	0.7686&0.3311&	0.7757&0.3401&0.7883&0.3502&	0.7918	\\
					0.5&0.3221&	0.7211&0.3431&	0.7379&0.3489&0.7365&0.3566&	0.7482	\\
					0.6&0.3301&	0.7520&0.3501&	0.7522&0.3555&0.7625&0.3602&	0.7629	\\
					0.7&0.2807&	0.6683&0.2889&	0.6694&0.2912&0.6715&0.3099&	0.6771 \\
					0.8&0.2908&  0.5487&0.3001&	0.5674&0.3092&0.5685&0.3112&	0.5775\\
					0.9&0.2011&	0.3906&0.2102&	0.4026&0.2199&0.4106&0.2245&	0.4120	\\
					\hline
					overall&0.2797&\textbf{0.6404} &0.2796&\textbf{0.6442}&0.2898&\textbf{0.6558}	&0.2948&\textbf{0.6572}\\
					\hline
			\end{tabular}}
			\scalebox{0.8}{
				\tabcolsep=0.4cm
				\begin{tabular}{lllllll}				
					\hline
					delta bucket&Fea6&\textbf{Fea6-BS}&Fea7&\textbf{Fea7-BS}&Fea3-CL&\textbf{Fea3-CL-BS} \\ 
					\hline
					0.1&0.3178&	0.6801&0.3738&	0.6822&0.3223&0.6771\\
					0.2&0.2589&	0.6875&0.2891&	0.6932&0.3712&0.6494\\
					0.3&0.2173&	0.6139&0.2224&	0.6048&0.4237&0.6481\\
					0.4&0.3612&	0.7994&0.3666&	0.8101&0.4189&0.7511\\
					0.5&0.3666&	0.7599&0.3701&	0.7610&0.4490&0.7482\\
					0.6&0.3733&	0.7711&0.3841&	0.7732&0.3998&0.6472\\
					0.7&0.3188&	0.6823&0.3223&	0.6925&0.3892&0.5485\\
					0.8&0.3232&	0.6083&0.3299&	0.6209&0.4112&0.4494\\
					0.9&0.2241&	0.4250&0.2356&	0.4287&0.2981&0.4551\\
					\hline
					overall &0.3068&\textbf{0.6697} &0.3215&\textbf{0.6741 }&0.3870& \textbf{0.6193}
					\\
					\hline
			\end{tabular}}
		\end{table}
	\end{center}
	\begin{table}[!htb]
		\caption{Gain ratios of daily hedging for S\&P 500 put options in various models}\label{tab-dayput}
		\scalebox{0.8}{
			\tabcolsep=0.4cm
			\begin{tabular}{lllllllll}				
				\hline
				delta bucket&Fea2&\textbf{Fea2-BS}&Fea3&\textbf{Fea3-BS}&Fea4&\textbf{Fea4-BS} &Fea5&\textbf{Fea5-BS}\\ 
				\hline			
				-0.1&0.1669&	0.2340&0.1416& 0.2468&0.1729&0.2522&0.1732&	0.2722\\
				-0.2&0.1587&	0.3766&0.1631& 0.2067&0.1643&0.2251&0.1747&	0.2439\\
				-0.3&0.1778&	0.3345&0.1799& 0.2880&0.1813&0.2103&0.1822&	0.3428\\
				-0.4&0.1546&	0.2273&0.1619& 0.2605&0.1646&0.2861&0.1662&	0.3993\\
				-0.5&0.1651&	0.3554&0.1666& 0.3734&0.1697&0.3859&0.1732&	0.3951\\
				-0.6&0.1588&	0.2017&0.1592& 0.3032&0.1603&0.4051&0.1653&	0.4172\\
				-0.7&0.1232&	0.3361&0.1311& 0.3416&0.1390&0.3671&0.1421&	0.3863\\
				-0.8&0.1334&	0.3166&0.1399& 0.3201&0.1467&0.3531&0.1499&	0.3666\\
				-0.9&0.1023&	0.3085&0.1165& 0.3112&0.1203&0.3322&0.1283&	0.3577\\
				\hline
				overall&0.1490&\textbf{0.2990}&0.1511&\textbf{0.2946}&0.1577&\textbf{0.3130}&0.1617&\textbf{0.3535}\\
				\hline
		\end{tabular}}
		\scalebox{0.8}{
			\tabcolsep=0.4cm
			\begin{tabular}{lllllll}				
				\hline
				delta bucket&Fea6&\textbf{Fea6-BS}&Fea7&\textbf{Fea7-BS} &Fea3-CL&\textbf{Fea3-CL-BS}\\ 
				\hline	
				-0.1&0.1811&	0.3838&0.1814   &0.3890&0.3382&0.3447\\
				-0.2&0.2015&    0.4670&0.2341	&0.4086&0.2123&0.3377\\
				-0.3&0.1927&	0.4642&0.1941	&0.4671&0.2311&0.4358\\
				-0.4&0.1689&	0.4577&0.1698	&0.4584&0.2001&0.4417\\
				-0.5&0.1789&	0.5078&0.1801	&0.5443&0.2019&0.4386\\
				-0.6&0.1715&	0.5322&0.1789   &0.5389&0.1982&0.4371\\
				-0.7&0.1479&	0.4112&0.1593	&0.4364&0.1132&0.4325\\
				-0.8&0.1588&	0.3922&0.1634	&0.4266&0.1338&0.4012\\
				-0.9&0.1342&	0.3998&0.1381	&0.4181&0.0983&0.3725\\
				\hline
			     overall&0.1706&\textbf{0.4462}&0.1777&\textbf{0.4542}&0.1919&\textbf{0.4046}\\
				\hline
		\end{tabular}}
	\end{table}
We have the following findings:

1. For each selected feature set, the gain ratio from learning the residual function is significantly higher than that from directly learning the hedging function, often with an improvement exceeding 100\%. For example, for the model group using TTM and $\delta_{BS}$ as features, the model Fea2-BS shows a 129\% improvement in the overall gain ratio for calls and a 101\% improvement for puts compared to the model Fea2. These improvements in the gain ratio are robust across different delta buckets.

2. For the models that learn the residual function, increasing the number of input features related to the option characteristics and their Greeks usually improves the gain ratio. For both calls and puts, the best-performing model is Fea7-BS. However, the improvement in the overall gain ratio from model Fea2 to Fea7 is only 5\% for calls, while for puts, the improvement is more significant, reaching 52\%. The implication is that, for call options, the model Fea2-BS with fewer input features is sufficient. However, for put options, it is necessary to add or select more appropriate input features. In such cases, models like Fea6-BS or Fea7-BS should be considered.

3. The models learning the residual function, with sufficient  option characteristics and their Greeks as input features, outperform the three-feature model that considers the market sentiment variable. For both calls and puts, the gain ratio of Fea6-BS or Fea7-BS is higher than that of Fea3-CL-BS. In contrast, for models that directly learn the hedging function, increasing the number of input features related to option characteristics and their Greeks does not significantly improve the gain ratio, and none of them surpass the three-feature model Fea3-CL, consistent with the findings in \cite{Chen2023}. These results indicate that when learning the residual function, the market sentiment variable is not as crucial as it is when directly learning the hedging function.

\begin{figure}[!htb]
		\centering
		\subfloat[]{\includegraphics[width=8cm, height=6cm]{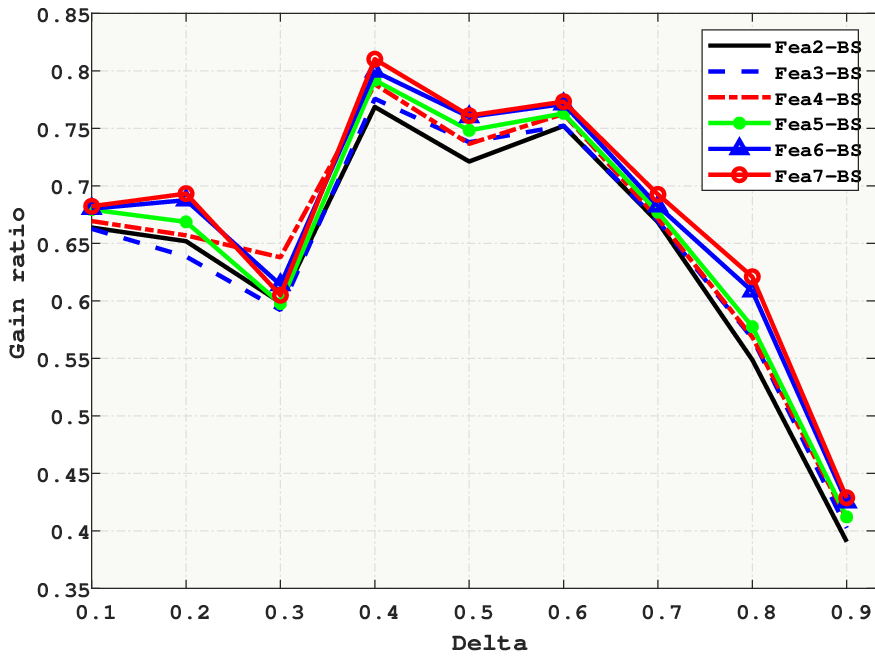}}
		\subfloat[]{\includegraphics[width=8cm, height=6cm]{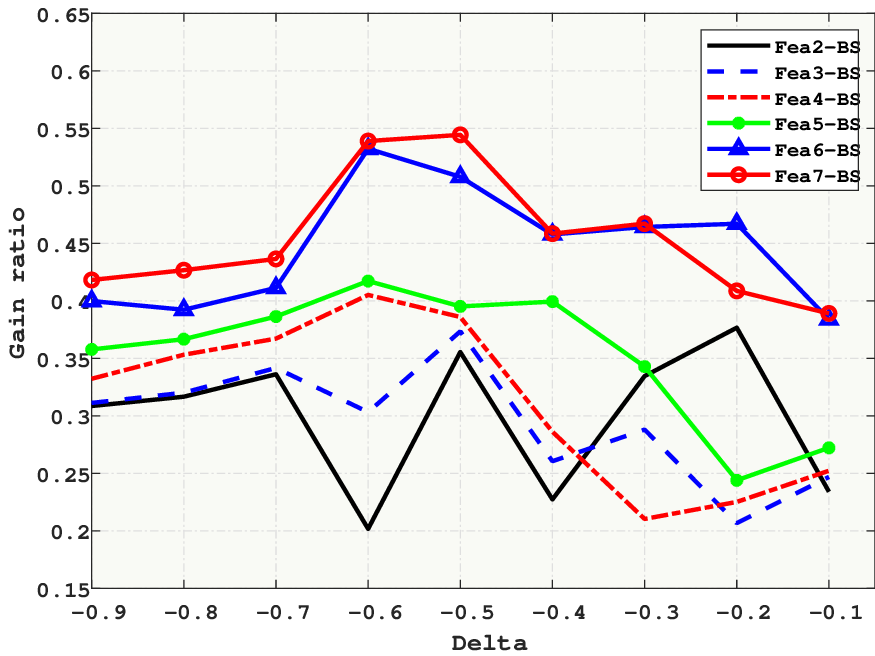}}
		\caption{\footnotesize (a) Gain ratios across delta buckets for call options under daily hedging. (b) Gain ratios across delta buckets for put options under daily hedging}
		\label{fig_delta}
	\end{figure}

4. For each delta bucket, the gain ratios are generally highest when the delta is closer to 0.5 for calls (and -0.5 for puts). At this point, the options are near the money, with the highest trading volume and best liquidity. \cref{fig_delta} plots the gain ratios across delta buckets for each model that learns the residual function. This figure clearly shows that for call options, the trend is consistent across all models. For put options, this result is more pronounced for models with a larger number of features, such as Fea6-BS and Fea7-BS.

\begin{figure}[!htb]
		\centering
		\subfloat[]{\includegraphics[width=8cm, height=6cm]{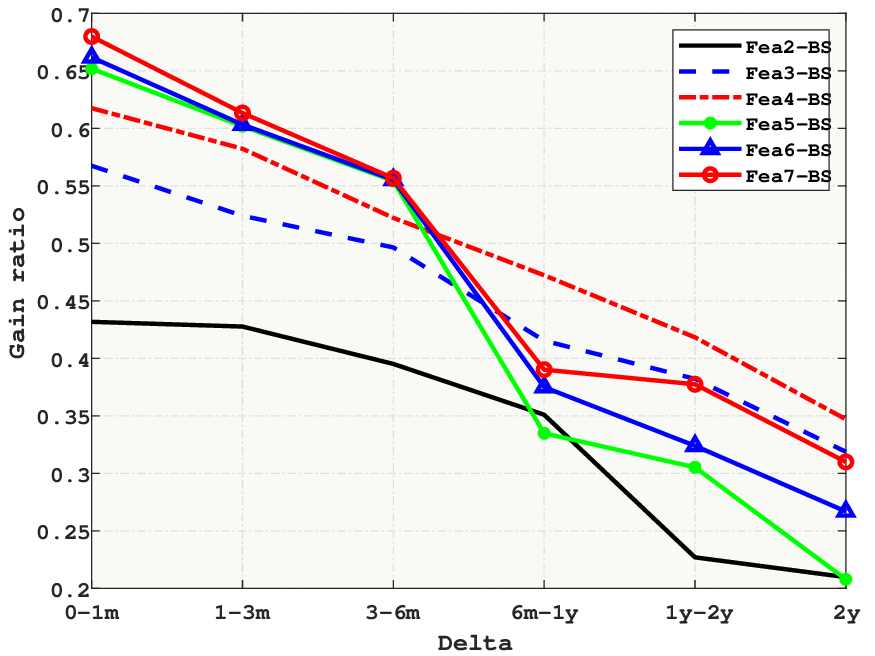}}
		\subfloat[]{\includegraphics[width=8cm, height=6cm]{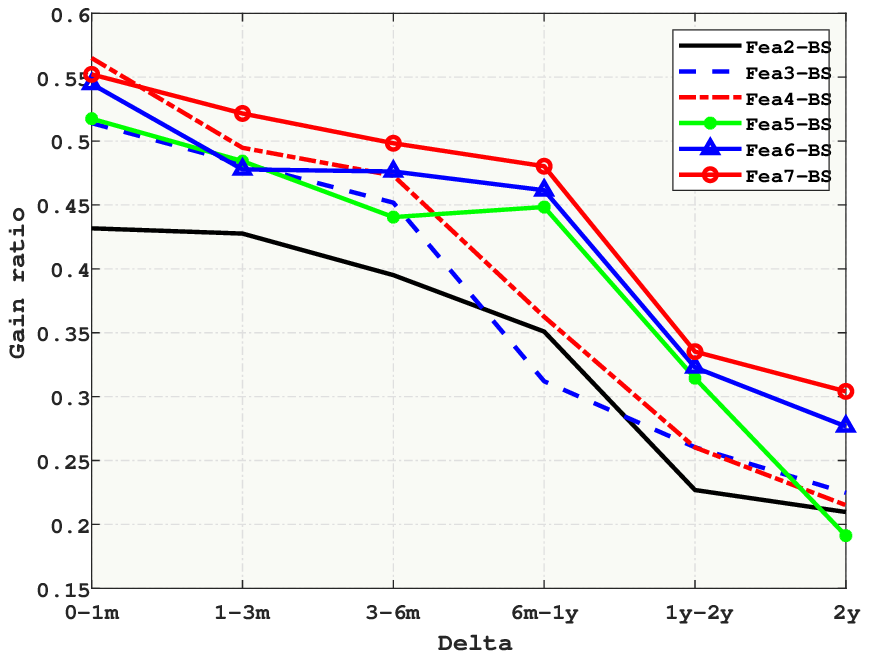}}
		\caption{\footnotesize (a) Gain ratios of daily hedging for S\&P 500 call options with different time to maturity in various models. (b) Gain ratios of daily hedging for S\&P 500 put options with different time to maturity in various models.}
		\label{fig_ttmdelta}
	\end{figure}

 We also divided the options in the test set into six categories based on their time to maturity: 0-1 month, 1-3 months, 3-6 months, 6 months-1 year, 1-2 years, and \textgreater 2 years. \cref{fig_ttmdelta} show the gain ratios across these six categories for each model that learns the residual function. The results show that the shorter the time to maturity, the higher the gain ratio. A plausible explanation is that options with shorter time to maturity have prices that are more sensitive to changes in the underlying asset price, making the deep hedging more effective in improving upon the implied Black-Scholes delta. Additionally, short-term options tend to have better liquidity.
	

\subsection{Weekly and monthly hedging performance of various models using 10-year data}

We also evaluated the hedging performance of learning the residual function using FNN under weekly and monthly hedging frequencies.  
\crefrange{tab-weekcall}{tab-monthput} show the gain ratios for weekly and monthly hedging, both for calls and puts, respectively. These tables indicate that for both weekly and monthly hedging: (1) The gain ratio from learning the residual function is significantly higher than that from directly learning the hedging function. (2) For the models that learn the residual function, increasing the number of input features related to the option characteristics and their Greeks usually improves the gain ratio. Different from daily hedging, for both calls and puts,  the improvements in the overall gain ratio from model Fea2 to Fea7 are significant. Thus, for weekly and monthly hedging, the Fea7-BS model is preferred for both call and put options. (3) When learning the residual function, for both calls and puts, the gain ratios of  Fea7-BS are higher than that of Fea3-CL-BS. Thus, the market sentiment variable is not as crucial as it is when directly learning the hedging function for weekly and monthly hedging. 

\begin{figure}[!htb]
		\centering
		\subfloat[]{\includegraphics[width=8cm, height=6cm]{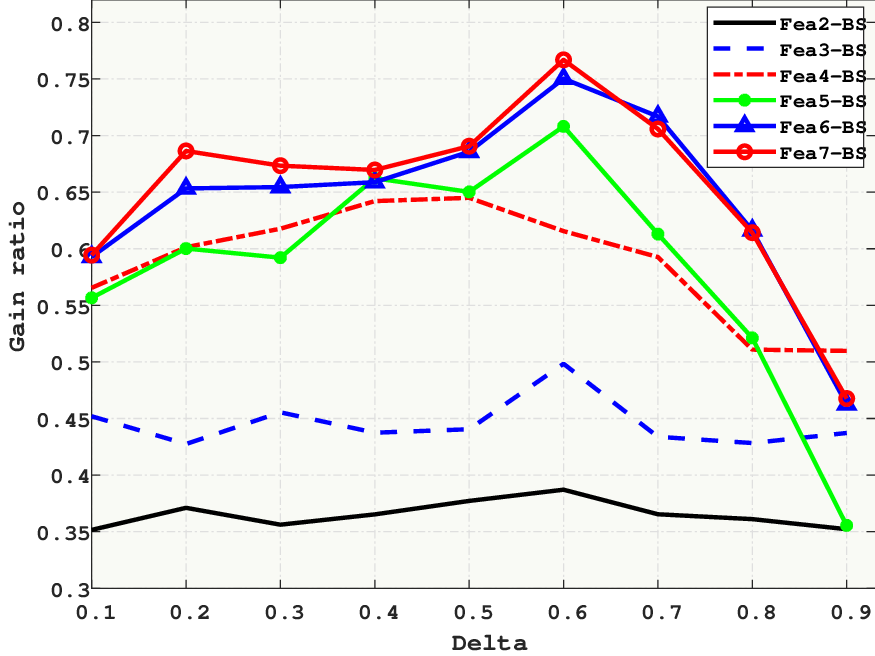}}
		\subfloat[]{\includegraphics[width=8cm, height=6cm]{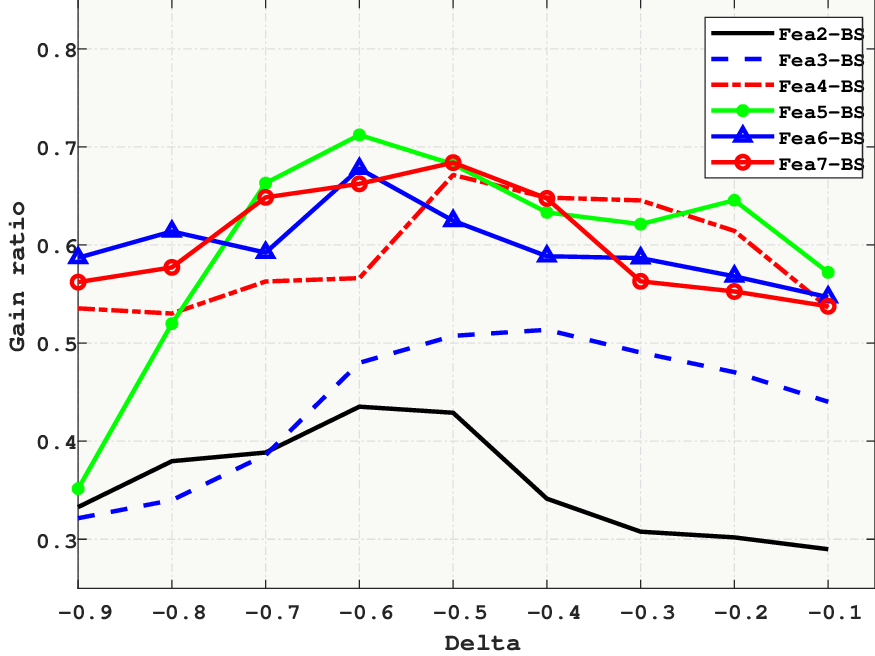}}
		
		\subfloat[]{\includegraphics[width=8cm, height=6cm]{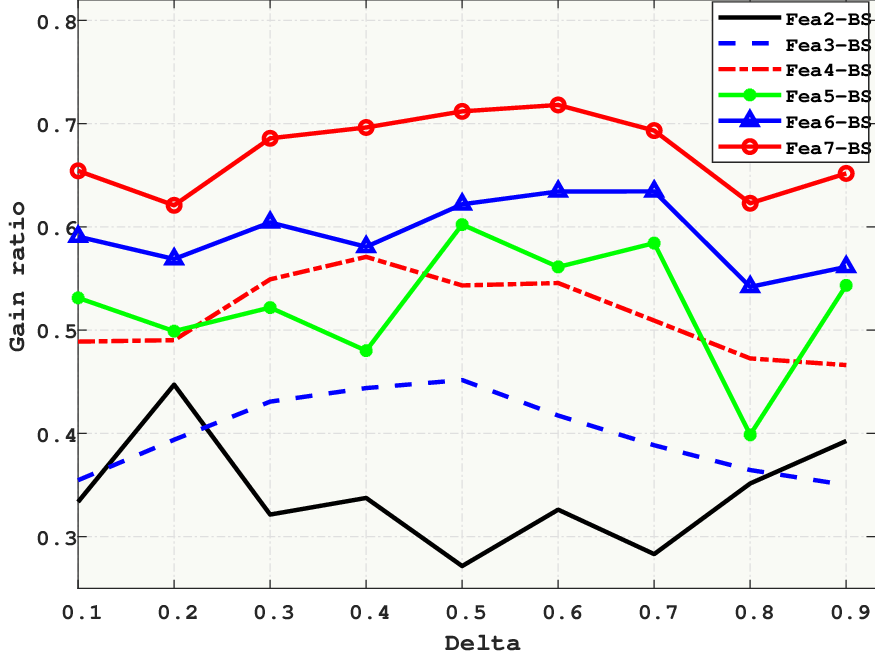}}
		\subfloat[]{\includegraphics[width=8cm, height=6cm]{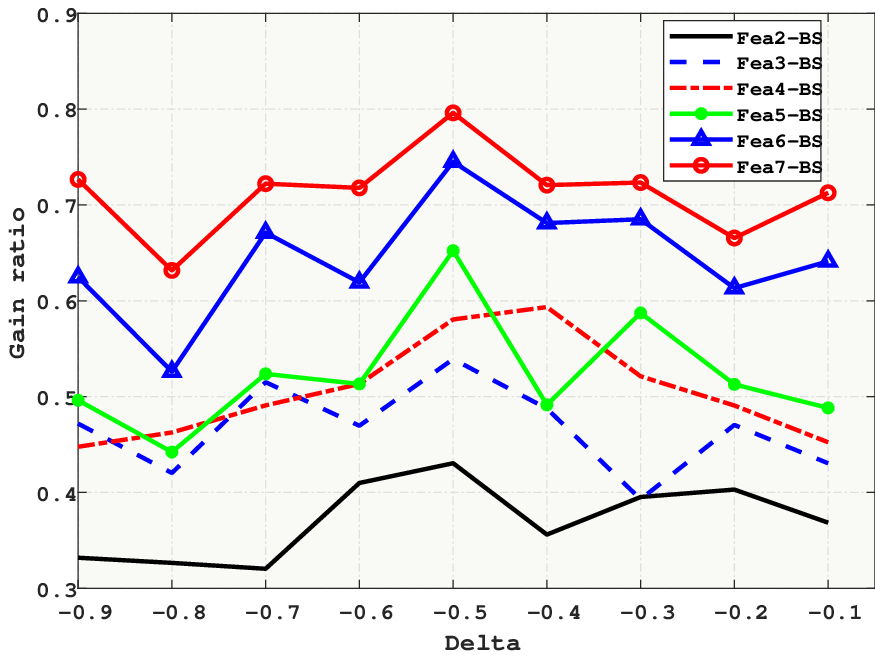}}
		\caption{\footnotesize (a) Gain ratios of call options under weekly hedging. (b) Gain ratios of put options under weekly hedging.
		(c) Gain ratios of call options under monthly hedging. (d) Gain ratios of put options under monthly hedging}
		\label{fig_wmdelta}
	\end{figure}

	\begin{table}[!htb]
	\caption{Gain ratios of weekly hedging for S\&P 500 call options in various models.}\label{tab-weekcall}
	\scalebox{0.8}{
		\tabcolsep=0.4cm
		\begin{tabular}{lllllllll}				
			\hline
			delta bucket&Fea2&\textbf{Fea2-BS}&Fea3&\textbf{Fea3-BS}&Fea4&\textbf{Fea4-BS} &Fea5&\textbf{Fea5-BS}\\ 
			\hline			
			0.1&0.1246&	0.3516&	0.2082&	0.4520&	0.1927 &0.5655&	0.2467&	0.5566	\\
			0.2&0.1411&	0.3711&	0.2374&	0.4275&	0.2049 	&0.6014&	0.2781&	0.6002\\
			0.3&0.1519&	0.3562&	0.1863&	0.4555&	0.2193 	&0.6177&	0.2899&	0.5921\\
			0.4&0.1857&	0.3653&	0.2788&	0.4375&	0.2522 	&0.6421&	0.2670&	0.6623\\
			0.5&0.1913&	0.3772&	0.2760&	0.4405&	0.2950 	&0.6450&	0.2833&	0.6502	\\
			0.6&0.2112&	0.3871&	0.2164&	0.4984&	0.2817 	&0.6156&	0.2230&	0.7081	\\
			0.7&0.2014&	0.3654&	0.2332&	0.4338&	0.2477 	&0.5927&	0.2313&	0.6130\\
			0.8&0.1927&	0.3611&	0.2043&	0.4283&	0.1965 	&0.5109&	0.2567&	0.5212	\\
			0.9&0.1884&	0.3522&	0.2005&	0.4372&	0.1871 	&0.5098&	0.1891&	0.3555	\\
			\hline
			overall&0.1765&\textbf{0.3652}&0.2268&\textbf{0.4456}&0.2308 &\textbf{0.5889}&0.2517&	\textbf{0.5844}	\\
			
			\hline
	\end{tabular}}
	\scalebox{0.8}{
		\tabcolsep=0.4cm
		\begin{tabular}{lllllll}				
			\hline
			delta bucket&Fea6&\textbf{Fea6-BS}&Fea7&\textbf{Fea7-BS}&Fea3-CL&\textbf{Fea3-CL-BS} \\ 
			\hline	
			0.1&	0.2551&	0.5928&	0.2601&	 0.5944&0.1688&0.4676\\
			0.2&	0.2822&	0.6533&	0.3101&	 0.6864&0.1917&0.6203\\
			0.3&	0.3102&	0.6545&	0.3322&	 0.6734&0.2187&0.6123\\
			0.4&	0.2591&	0.6588&	0.2720&	 0.6696&0.2431&0.6413\\
			0.5&	0.2898&	0.6857&	0.2978&	 0.6906&0.2159&0.6734\\
			0.6&	0.2101&	0.7503&	0.2001&	 0.7669&0.2346&0.6276\\
			0.7&	0.2011&	0.7169&	0.1909&	 0.7058&0.2178&0.5331\\
			0.8&	0.2101&	0.6162&	0.2782&	 0.6142&0.1851&0.4821\\
			0.9&	0.1802&	0.4625&	0.1967&	 0.4677&0.1438&0.3753\\
			\hline
			overall&	0.2442&	\textbf{0.6434}&	0.2598&\textbf{0.6521}&0.2022&\textbf{0.5592}\\
			\hline
	\end{tabular}}
\end{table}
	\begin{table}[!htb]
	\caption{Gain ratios of weekly hedging for S\&P 500 put options in various models.}\label{tab-weekput}
	\scalebox{0.8}{
		\tabcolsep=0.4cm
		\begin{tabular}{lllllllll}				
			\hline
			delta bucket&Fea2&\textbf{Fea2-BS}&Fea3&\textbf{Fea3-BS}&Fea4&\textbf{Fea4-BS} &Fea5&\textbf{Fea5-BS}\\ 
			\hline			
			-0.1&	0.2286& 	0.2898 &	0.1792 &	0.4401 &	0.1830 &0.5363&	0.1912&	0.5721\\
			-0.2&	0.2206& 	0.3019 &	0.1982 &	0.4703 &	0.1941 &0.6143&	0.2054&	0.6456\\
			-0.3&	0.1507&		0.3077 &	0.1883 &	0.4903 &	0.2350 &0.6454&	0.1883&	0.6212\\
			-0.4&	0.2380& 	0.3414 &	0.1715 &	0.5135 &	0.2135 &0.6483&	0.1981&	0.6332\\
			-0.5&	0.2671& 	0.4290 &	0.2858 &	0.5074 &	0.2714 &0.6714&	0.2309&	0.6823 \\
			-0.6&	0.2170& 	0.4351 &	0.2957& 	0.4799 &	0.2844 &0.5662&	0.2201&	0.7122 \\
			-0.7&	0.2088& 	0.3884 &	0.2127& 	0.3859 &	0.1899 &0.5628&	0.1876&	0.6631\\
			-0.8&	0.1846& 	0.3796 &	0.2096& 	0.3398 &	0.1777 &0.5301&	0.2289&	0.5198\\
			-0.9&	0.1934& 	0.3328 &	0.1589& 	0.3214 &	0.1548 &0.5354&	0.2333&	0.3515\\
			\hline
			overall&0.2121 &	\textbf{0.3562} &0.2111 &\textbf{0.4387}&	0.2115 &\textbf{0.5274}&0.2093&\textbf{0.6001}\\
			
			\hline
	\end{tabular}}
	\scalebox{0.8}{
		\tabcolsep=0.4cm
		\begin{tabular}{lllllll}				
			\hline
			delta bucket&Fea6&\textbf{Fea6-BS}&Fea7&\textbf{Fea7-BS}&Fea3-CL&\textbf{Fea3-CL-BS} \\ 
			\hline			
			-0.1&	0.2012&0.5468	&	0.2131&0.5375&0.2423&0.4896	\\
			-0.2&	0.2145&0.5680	&	0.2236&0.5526&0.3981&0.5282	\\
			-0.3&	0.1902&0.5866	&	0.2125&0.5629&0.2223&0.5831	\\
			-0.4&	0.2253&0.5885	&	0.2195&0.6475&0.3318&0.5844\\
			-0.5&	0.2275&0.6247	&	0.2389&0.6840&0.3721&0.5881	\\
			-0.6&	0.2332&0.6778	&	0.2413&0.6623&0.2591&0.6155	\\
			-0.7&	0.2021&0.5922	&	0.2322&0.6485&0.1989&0.6802	\\
			-0.8&	0.2301&0.6138	&	0.2431&0.5771&0.3217&0.6449	\\
			-0.9&	0.2412&0.5869	&	0.2531&0.5621&0.2191&0.5422	\\
			\hline
			overall&0.2184&	\textbf{0.5984}&0.2308&\textbf{0.6038}&0.2850&\textbf{0.5840}\\
			\hline
	\end{tabular}}
\end{table}
	\begin{table}[!htb]
	\caption{Gain ratios of monthly hedging for S\&P 500 call options in various models.}\label{tab-monthcall}
	\scalebox{0.8}{
		\tabcolsep=0.4cm
		\begin{tabular}{lllllllll}				
			\hline
			delta bucket&Fea2&\textbf{Fea2-BS}&Fea3&\textbf{Fea3-BS}&Fea4&\textbf{Fea4-BS}&Fea5&\textbf{Fea5-BS} \\ 
			\hline			
			0.1&	0.1796& 	0.3337& 	0.1873& 	0.3547& 	0.2442 &0.4889&	0.1199&	0.5313\\
			0.2&	0.1918& 	0.4472& 	0.2078& 	0.3939& 	0.2090 &0.4903&	0.1566&	0.4990\\
			0.3&	0.1851& 	0.3214& 	0.2381& 	0.4308& 	0.2275 &0.5492&	0.1540&	0.5219\\
			0.4&	0.2572& 	0.3375& 	0.2590& 	0.4439& 	0.2341 &0.5710&	0.1448&	0.4801\\
			0.5&	0.2734& 	0.2715& 	0.2833& 	0.4517& 	0.2585 &0.5433&	0.1481&	0.6023 \\
			0.6&	0.2486& 	0.3261& 	0.2357& 	0.4173& 	0.2238 &0.5456&	0.1509&	0.5612\\
			0.7&	0.1739& 	0.2831& 	0.2159& 	0.3885& 	0.2706 &0.5093&	0.1633&	0.5843\\
			0.8&	0.1688& 	0.3515& 	0.2207& 	0.3645& 	0.2011 &0.4726&	0.1679&	0.3987\\
			0.9&	0.1485& 	0.3927& 	0.1711& 	0.3500& 	0.1976 &0.4662&	0.1578&	0.5434\\
			\hline
			overall&0.2030 &	\textbf{0.3405} &0.2243& \textbf{0.3995}& 0.2296 &\textbf{0.5152}&0.1515&	\textbf{0.5247}	\\
			
			\hline
	\end{tabular}}
	\scalebox{0.8}{
		\tabcolsep=0.4cm
		\begin{tabular}{lllllll}				
			\hline
			delta bucket&Fea6&\textbf{Fea6-BS}&Fea7&\textbf{Fea7-BS}&Fea3-CL&\textbf{Fea3-CL-BS} \\ 
			\hline			
			0.1	&	0.1317&	0.5906&	0.1624&	0.6543&	0.2013&0.5202\\
			0.2	&	0.1654&	0.5689&	0.1678&	0.6209&	0.2267&0.4776\\
			0.3	&	0.1602&	0.6043&	0.1717&	0.6858&	0.1897&0.5163\\
			0.4	&	0.1479&	0.5807&	0.1569&	0.6963&	0.1799&0.4935\\
			0.5	&	0.1523&	0.6219&	0.1661&	0.7119&	0.1981&0.5363\\
			0.6	&	0.1621&	0.6343&	0.1779&	0.7181&	0.2321&0.5421\\
			0.7	&	0.1709&	0.6344&	0.1890&	0.6933&	0.2410&0.5411\\
			0.8	&	0.1898&	0.5419&	0.1977&	0.6229&	0.1729&0.4513\\
			0.9	&	0.1623&	0.5613&	0.1802&	0.6518&	0.1601&0.4731\\
			\hline
			overall&	0.1603&	\textbf{0.5931}&	0.1744&\textbf{0.6728}&0.2002&\textbf{0.5057}\\
			\hline
	\end{tabular}}
\end{table}
	\begin{table}[!htb]
	\caption{Gain ratios of monthly hedging for S\&P 500 put options in various models.}\label{tab-monthput}
	\scalebox{0.8}{
		\tabcolsep=0.4cm
		\begin{tabular}{lllllllll}				
			\hline
			delta bucket&Fea2&\textbf{Fea2-BS}&Fea3&\textbf{Fea3-BS}&Fea4&\textbf{Fea4-BS}&Fea5&\textbf{Fea5-BS} \\ 
			\hline			
			-0.1&0.2524& 	0.3685& 	0.2641& 	0.4305& 	0.1971  &0.4523&	0.3153&	0.4882	\\
			-0.2&0.2330&	0.4030& 	0.2737& 	0.4705& 	0.1955 	&0.4907&	0.3001&	0.5128\\
			-0.3&0.3254& 	0.3953& 	0.2955& 	0.3929& 	0.2511 	&0.5212&	0.3204&	0.5873\\
			-0.4&0.3532& 	0.3560& 	0.3288& 	0.4876& 	0.2563 	&0.5934&	0.3617&	0.4914\\
			-0.5&0.3785& 	0.4305& 	0.2872& 	0.5392& 	0.2207 	&0.5805&	0.3303&	0.6523\\
			-0.6&0.2026& 	0.4099& 	0.2554& 	0.4696& 	0.2174 	&0.5132&	0.3217&	0.5132\\
			-0.7&0.1936& 	0.3204& 	0.2195& 	0.5151& 	0.2175 	&0.4909&	0.2891&	0.5237\\
			-0.8&0.1814& 	0.3265& 	0.2061& 	0.4203& 	0.1864 	&0.4626&	0.2717&	0.4422\\
			-0.9&0.1621& 	0.3319& 	0.1852& 	0.4718& 	0.1634 	&0.4477&	0.2563&	0.4962\\
			\hline
			overall&0.2536 &	\textbf{0.3713} &	0.2573 &	\textbf{0.4664 }&	0.2117 &\textbf{0.5058}&0.3074&	\textbf{0.5230}	\\
			
			\hline
	\end{tabular}}
	\scalebox{0.8}{
		\tabcolsep=0.4cm
		\begin{tabular}{lllllll}				
			\hline
			delta bucket&Fea6&\textbf{Fea6-BS}&Fea7&\textbf{Fea7-BS}&Fea3-CL&\textbf{Fea3-CL-BS} \\ 
			\hline			
			-0.1&	0.3213&	0.6412&	0.3224&	0.7127&	0.2187&0.5609\\
			-0.2&	0.3067&	0.6132&	0.3135&	0.6655&	0.2289&0.6276\\
			-0.3&	0.3324&	0.6851&	0.3512&	0.7234&	0.2364&0.6036\\
			-0.4&	0.3773&	0.6811&	0.4263&	0.7207&	0.1987&0.6398\\
			-0.5&	0.3211&	0.7451&	0.3543&	0.7961&	0.2076&0.6647\\
			-0.6&	0.3431&	0.6192&	0.3613&	0.7178&	0.1881&0.7198\\
			-0.7&	0.3128&	0.6711&	0.3221&	0.7222&	0.2354&0.6816\\
			-0.8&	0.2822&	0.5261&	0.2902&	0.6317&	0.1879&0.6768\\
			-0.9&	0.2671&	0.6247&	0.2712&	0.7266&	0.1885&0.5178\\
			\hline
			overall&	0.3182&	\textbf{0.6452}&	0.3347&\textbf{0.7130}&0.2100&\textbf{0.6325}	\\
			\hline
	\end{tabular}}
\end{table}
\cref{fig_wmdelta} plots the gain ratios for weekly and monthly hedging, both for calls and puts,  across delta buckets. This figure also shows that the gain ratios are generally highest when the delta is closer to 0.5 for calls (and -0.5 for puts) for weekly and monthly hedging, especially for models with more input features, such as  Fea6-BS and Fea7-BS. These empirical results demonstrate the robustness of our method in improving hedging effectiveness across different hedging frequencies.   

   \subsection{Hedging performance of various models using 3-year subdatasets}

In this subsection, we consider learning hedging strategies that minimize the mean squared local hedging error using only three years of data. Empirical results in this subsection show that learning the residual function with three years of data can achieve comparable hedging performance to directly learning the hedging function with ten years of data. This further demonstrates that the smoother nature of the residual function makes it easier for neural networks to approximate and requires less data for training.

Specifically, we divide the nine years of data from January 1, 2010, to December 31, 2018, into three segments: 2010-2012, 2013-2015, and 2016-2018. For each three-year segment, we use 
the first two years  of data for training and validation and the last year for testing. 
We employed two hidden layers with 128 neurons in each hidden layer, which is relatively simpler compared to the neural network handling ten years of data. We set the learning rate to 0.0001. In our experiments, training typically stopped after about 40 epochs.

We present the empirical results for daily hedging. Based on the results from using ten years of data, we use the model Fea2/Fea2-BS to evaluate the hedging performance with three years of data for call options (see \cref{tab-3yearcall-Fea2}) and the model Fea7/Fea7-BS for put options (see \cref{tab-3yearput-Fea7}).
   
   \begin{center}
		\begin{table}[!htb]
			\caption{Gain ratios of daily hedging for S\&P 500 call options in Fea-2/Fea-2-BS models using different three-year data.}\label{tab-3yearcall-Fea2}
			\scalebox{0.8}{
				\tabcolsep=0.55cm
				\begin{tabular}{llllllll}				
					\hline
					delta bucket&\multicolumn{2}{c}{2010-2012}&\multicolumn{2}{c}{2013-2015}&\multicolumn{2}{c}{2016-2018}\\ 
 \hline    
                    &Fea-2&Fea-2-BS&Fea-2&Fea-2-BS&Fea-2&Fea-2-BS\\
					\hline
					0.1&0.0591&	0.3443&	0.0552&	0.2913&	0.0465&	0.2719	\\
					0.2&0.1114&	0.3964&	0.1180&	0.3581&	0.1051&	0.3731	\\
					0.3&0.1543&	0.4054&	0.1567&	0.3675&	0.1843&	0.3902	\\
					0.4&0.1798&	0.4105&	0.1630&	0.3716&	0.1876&	0.4004	\\
					0.5&0.1875&	0.4149&	0.1721&	0.3745&	0.1957&	0.4103	\\
					0.6&0.1442&	0.4112&	0.1746&	0.3649&	0.1499&	0.3983	\\
					0.7&0.0536&	0.4066&	0.0883&	0.3430&	0.0992&	0.3797 \\
					0.8&0.0187&	0.4046&	0.0263&	0.3283&	0.0388&	0.3628\\
					0.9&0.0173&	0.3847&	-0.0636&0.3004&	-0.0545&0.2298	\\
					\hline
					overall&0.1029& \textbf{0.3976}&0.0990&\textbf{0.3444} &0.1058&\textbf{0.3574}	\\
					\hline
			\end{tabular}}
		\end{table}
        \begin{table}[!htb]
	\caption{Gain ratios of daily hedging for S\&P 500 put options in Fea-7/Fea-7-BS models using different three-year data.}\label{tab-3yearput-Fea7}
	\scalebox{0.8}{
		\tabcolsep=0.55cm
		\begin{tabular}{lllllll}				
			\hline
			delta bucket&\multicolumn{2}{c}{2010-2012}&\multicolumn{2}{c}{2013-2015}&\multicolumn{2}{c}{2016-2018}\\ 
			\hline
			&Fea-7&\textbf{Fea-7-BS}&Fea-7&\textbf{Fea-7-BS}&Fea-7&\textbf{Fea-7-BS}\\
			\hline
			-0.1	&0.0960&0.3413&	0.1232&	0.3872&	0.1138&	0.3608\\
			-0.2	&0.1060&0.3493&	0.1177&	0.4082&	0.1376&	0.3510\\
			-0.3	&0.1395&0.3764&	0.1418&	0.4374&	0.1370&	0.3493\\
			-0.4	&0.1138&0.4069&	0.1699&	0.4386&	0.1801&	0.3698\\
			-0.5	&0.1203&0.4385&	0.1812&	0.4179&	0.1797&	0.4052\\
			-0.6	&0.1870&0.4151&	0.1581&	0.3760&	0.1407&	0.4282\\
			-0.7	&0.1711&0.3900&	0.1484&	0.3638&	0.0836&	0.4057\\
			-0.8	&0.1650&0.3535&	0.1318&	0.3433&	0.0779&	0.3755\\
			-0.9	&0.1313&0.3512&	0.1287&	0.3404&	0.0701&	0.3377\\
			\hline
			overall&0.1367&	\textbf{0.3802}&0.1445&	\textbf{0.3903}	&0.1245	&\textbf{0.3759}	\\
			\hline
	\end{tabular}}
\end{table}
   \end{center}

Comparing \cref{tab-3yearcall-Fea2} and \cref{tab-daycall}, we observe that for the periods 2010-2012, 2013-2015, and 2016-2018 using three-year data, the overall gain ratios of the Fea2-BS model (learning the residual function) are 0.3976, 0.3444, and 0.3574, respectively. These ratios exceed the gain ratio of 0.2797 obtained under the Fea2 model and are comparable to the gain ratio of 0.3870 obtained under the Fea3-CL model, both using the ten-year data and directly learning the hedging function. 
In contrast, if directly learning the hedging function using three years of data, the overall gain ratios using the Fea-2 model are only 0.1029, 0.0990, and 0.1058 for the periods 2010-2012, 2013-2015, and 2016-2018. These hedging performances are insufficient, consistent with the findings of \cite{Chen2023}, which emphasize the need for larger datasets, such as 10 years, to directly learn the hedging function effectively. Similar conclusions can be drawn for put options by comparing \cref{tab-3yearput-Fea7} and \cref{tab-dayput}. Therefore, our results demonstrate that learning residuals can reduce the data requirements.

It is also worth mentioning that these gain ratios from learning the residual function using three-year data are not worse than those obtained in \cite{Nian2021} (see their Tables 5 and 6 for daily hedging), which developed a complex GRU model using three years of data integrating both local and time-sequential features. Therefore, leveraging the smoother functional properties of the residual function allows us to achieve effective hedging results even with the use of a simple neural network.

\section{Conclusions}\label{sec5}

This paper follows the literature in using deep learning for a data-driven approach to address the option delta hedging problem. We integrate the economic structure of the problem to enhance the effectiveness of deep learning: reducing data requirements and improving out-of-sample performance. Specifically, recognizing the non-smoothness issues in the hedging function and the economic structure of the implied Black-Scholes delta, we employ a classic feedforward neural network (FNN) to learn the residuals between the hedging function and the implied Black-Scholes delta, aiming to minimize mean squared local hedging errors. Through comprehensive analysis of ten years of S\&P 500 index option data, we demonstrate the effectiveness and robustness  of our approach. Compared to directly learning the hedging function, our approach significantly improves hedging performance. Even with relatively less data, our approach achieves acceptable enhancements over the hedging strategy using the implied Black-Scholes delta.

\section*{Funding}
This research was supported by  the National Natural Science Foundation of China (NSFC)  
under grants 72271157, 72171109, 72310107002 and 71972131. 

\section*{Data availability}
The authors do not have permission to share data.

\section*{Disclosure statement}

No potential conflict of interest was reported by the author(s).

	\bibliographystyle{elsarticle-harv}

	\bibliography{bmyref.bib}

\end{document}